\documentstyle[aaspptwo,psfig]{article}

\def\etal{{\it et al. }}

\begin{document}

\title{Globular Cluster Luminosity Functions and the Hubble Constant
from WFPC2 Imaging: Galaxies in the Coma I Cloud\altaffilmark{1}}

\author{Duncan A. Forbes}
\affil{Lick Observatory, University of California, Santa Cruz, CA
95064}
\affil{and}
\affil{School of Physics and Space Research, University of Birmingham,
Edgbaston, Birmingham B15 2TT, United Kingdom}
\affil{Electronic mail: forbes@lick.ucsc.edu}

\altaffiltext{1}{Based on observations with the NASA/ESA {\it Hubble
Space Telescope}, obtained at the Space Telescope Science Institute,
which is operated by AURA, Inc., under NASA contract NAS 5--26555}

\begin{abstract}

The membership of some galaxies in the nearby (d $\sim 12$ Mpc) Coma I cloud is
uncertain. Here we present globular cluster luminosity functions (GCLFs) from
the {\it Hubble Space Telescope} for two bright ellipticals which may belong to
this group.  After fitting the GCLF, we find a turnover magnitude of $m_V ^0$ =
23.23 $\pm$ 0.11 for NGC 4278 and $m_V ^0$ = 23.07 $\pm$ 0.13 for NGC 4494. 
Our limiting magnitude is about two magnitudes fainter than these values,
making this data among the most complete GCLFs published to date.  The fitted
GCLF dispersions ($\sim$ 1.1$^m$) are somewhat smaller than typical values for
other ellipticals.  Assuming an absolute turnover magnitude of M$_V^0$ =
--7.62, and after applying a small metallicity correction, we derive distance
modulii of (m -- M) = 30.61 $\pm$ 0.14 for NGC 4278 and 30.50 $\pm$ 0.15 for
NGC 4494.  These distance estimates are compared to other methods, and lie
within the published range of values. We conclude that both galaxies lie at the
same distance and are both members of the Coma I cloud.  

\end{abstract}


\section{Introduction}

Although galaxies cover a wide range in properties, such as luminosity, Hubble
type and environment, their globular cluster systems are remarkably similar in
many respects. In particular, all globular cluster luminosity functions (GCLF)
can be crudely represented by a Gaussian with a peak located at M$_V$ $\sim$
--7.5 and a width of $\sigma$ $\sim$ 1.2. The peak, or turnover magnitude, has
been used with some success as an independent distance estimator to galaxies
(see Jacoby \etal 1992). Its usefulness relies on the assumption that the
turnover magnitude is universal for all galaxies.  It has been suggested that
the GCLF is not universal, but rather varies systematically with GC metallicity
(Ashman, Conti \& Zepf 1995), the galaxy Hubble type (Secker \& Harris 1993) or
galaxy environment (Blakeslee \& Tonry 1996). If any of these `second
parameter' effects are confirmed, then accurate distance estimates using the
GCLF method will require an appropriate correction. In the case of GC
metallicity, Ashman \etal (1995) have quantified the effects of metallicity on
the turnover magnitude and show that once a metallicity correction ($\sim$ 0.1
mag) is applied, the GCLF gives distances in good agreement with other methods.

To fully resolve these issues, a number of high quality GCLFs are required
covering a range of Hubble types and environments. One approach is to study
galaxies that belong to a nearby group.  This has the benefit of allowing the
GCLF--determined distances to be directly compared with other distance methods
for galaxies in the same group.  Fleming \etal (1995) aimed to use this
approach on the nearby Coma I cloud by examining the GCLF in NGC 4494 (E1) and
NGC 4565 (Sb). Their results will be discussed below.  The Coma I cloud is a
small group of galaxies in the foreground of the well--known Coma cluster of
galaxies (A1656).  It is dominated by spiral galaxies but also includes a few
ellipticals (Gregory \& Thompson 1977).  

Here we examine the GCLFs of the two brightest ellipticals which may be members
of the Coma I cloud, namely NGC 4278 and NGC 4494.  Both galaxies contain a
kinematically--distinct core which may have resulted from a merger.  Using {\it
Hubble Space Telescope} Wide Field and Planetary Camera 2 (WFPC2) data of
Forbes \etal (1996), we can probe the GCLF in these galaxies to about 2
magnitudes beyond the turnover magnitude. The WFPC2 data have the additional
advantages of very low background contamination with no blending of GCs in the
central galaxy region. Our method is similar to that applied to the GCLF of NGC
4365 (Forbes 1996). In particular, we use the maximum likelihood code of Secker
\& Harris (1993) to fit the GCLF and the absolute turnover magnitude
determination of Sandage \& Tammann (1995), i.e. M$_V^0$ = --7.62 $\pm$ 0.08. 
We compare the GCLF distance modulus with other methods and estimate the Hubble
constant.  

\section{Observations and Data Reduction}

Details of the WFPC2 data for the GCs in NGC 4278 and NGC 4494 are presented,
along with 12 other ellipticals, in Forbes \etal (1996). From the 2 $\times$
500s F555W images, we use DAOPHOT (Stetson 1987) to detect GCs. Detection is
based on fairly conservative limits of flux threshold, shape, sharpness and
size. And after checking against a list of hot pixels, the final contamination
from cosmic rays, foreground stars, background galaxies and hot pixels is less
than a few percent. We did not apply any color selection to the GC lists. The
GC magnitudes have been converted into Johnson V and corrected for Galactic
extinction.  

When examining the GCLF of NGC 4365 (Forbes 1996), we showed that the fraction
of detected GCs with magnitude, once the threshold criterion is adjusted for a
0.3 mag sensitivity difference (Burrows \etal 1993), is similar for all four
CCDs.  For NGC 4494 we were able to use all CCDs, as the dust is confined to a
small ($\sim$ 1$^{''}$) ring which doesn't appear to affect any GCs (Carollo
\etal 1996). In the case of NGC 4278, dust covers much of the PC CCD. We have
therefore decided not to include GCs from the PC in this analysis.  

\section{Modeling}

A full description of the modeling processes including completeness tests,
photometric errors, background contamination and maximum likelihood fitting,
are given in Forbes (1996). To summarize, we carried out simulations to test
the ability of DAOPHOT to detect GCs on actual WFC images for both galaxies.
These completeness tests, and the photometric error in measuring GC magnitudes,
are given in Figure 1 for NGC 4278 and Figure 2 for NGC 4494. Both figures are
similar to that found for NGC 4365 (Forbes 1996). The 50\% completeness level
is V = 24.9 for NGC 4278 and V = 24.8 for NGC 4494. We ignore GCs with
magnitudes fainter than these to avoid large incompleteness corrections. A
small correction is made for background contamination based on similar exposure
WFPC2 images from the Medium Deep Survey (Forbes \etal 1994). Finally we fit
the background--corrected GCLF using the maximum likelihood code of Secker \&
Harris (1993) which takes account of the completeness and photometric error
variations with magnitude. We fit both Gaussian and $t_5$ distributions to the
GCLF.  

\section{Results and Discussion}

We have detected 241 GCs in NGC 4278 and 148 in NGC 4494 to the 50\%
completeness level.  The results of fitting the GCLF of each galaxy, with the
Gaussian and a $t_5$ functions, are summarized in Table 1. In particular, the
average turnover magnitudes are m$_V ^0$ = 23.23 $\pm$ 0.11 for NGC 4278 and
m$_V ^0$ = 23.07 $\pm$ 0.13 for NGC 4494.  As these magnitudes are
statistically within the combined errors, this would indicate that both
galaxies are at the same distance.  The probability contours output from the
maximum likelihood code for the Gaussian fit, over a range of 0.5--3 standard
deviations, are shown in Figures 3 and 4.  In Figures 5 and 6 we show the
binned GCLF and our best--fit Gaussian superposed for each galaxy.  Note that
the fitting procedure does not use binned data but rather treats each data
point individually. These figures clearly show that the 50\% completeness limit
is almost 2 magnitudes fainter than the turnover magnitude, giving us
additional confidence that the turnover is well determined.  This makes our
data among the most complete, in terms of sampling the luminosity function, of
GCLFs published to date. The number of `missing' GCs, i.e. those fainter than
the limiting magnitude, is $\le$ 10\%.  

Fleming \etal (1995) have recently investigated the GCLF of NGC 4494 and the Sb
spiral NGC 4565. The aim of their study was to derive a GCLF distance for
galaxies of different Hubble types located in the same group. This would allow
a direct test of whether or not the GCLF turnover magnitude depends on Hubble
type. They used the same CFHT $\sim$1$^{''}$ V band images used by Simard \&
Pritchet (1994) for a surface brightness fluctuation (SBF) study of these
galaxies.  The data consisted of two pointings for each galaxy.  Using DAOPHOT
for GC detection and completeness tests, their 50\% completeness limits for the
two pointings were V = 23.2, 24.0 for NGC 4494 and V = 24.5, 24.3 for NGC 4565.
For a Gaussian fit to their GCLF data of NGC 4494 they quote $m_V ^0$ = 23.6
$\pm$ 0.4 with $\sigma$ fixed at 1.4. The results for NGC 4565 are $m_V ^0$ =
22.63 $\pm$ 0.2, $\sigma$ = 1.35 $\pm$ 0.22. In both cases there is
considerable scatter in their faint magnitude bins (see their Figures 5 and 6).
They concluded that NGC 4565 was in the Coma I cloud but that NGC 4494 lies in
the background.  

We now consider whether their data, and hence inferred distances, are
consistent with our dataset. We first note that all Fleming \etal magnitudes
should be 0.05$^m$ brighter after applying a Galactic extinction correction. 
Starting with NGC 4494, for which their 50\% completeness limit is comparable
to their estimated turnover magnitude. For a Gaussian fit our data gives $m_V
^0$ = 23.05 $\pm$ 0.13 which is {\it not} consistent with their result.
However, an important distinction is that they fixed $\sigma$ to be 1.4 (their
data did not warrant fits to both $\sigma$ and $m_V ^0$). It has been shown
that uncertainties in $\sigma$ correspond to uncertainties in $m_V ^0$, so that
a change from $\sigma$ = 1.4 (Fleming \etal) to 1.1 (us) would translate into a
brighter $m_V ^0$ by about 0.5$^m$ (e.g. Hanes 1977; Secker \& Harris 1993).
Thus the Fleming \etal GCLF would have a turnover of $m_V ^0$ $\sim$ 23.05. An
alternative, and perhaps more straight forward, way to show this is given by
Figure 7. This shows the Fleming \etal data for NGC 4494 and the best fit
Gaussian to our data for NGC 4494. We include two Gaussians arbitrarily scaled
up and down by a factor of two vertically (the turnover magnitude is held
constant, but the dispersion is allowed to vary). This figure indicates that,
within the error bars, the Fleming \etal data for NGC 4494 are consistent with
a Gaussian that has a turnover of $m_V ^0$ = 23.05.  As a further test, we
re--fit their data but excluded the faintest bin and allowed the Gaussian
dispersion $\sigma$ to be a free parameter (along with the turnover magnitude
and the normalization). This gave m$_V ^0$ = 23.27 $\pm$ 0.4 (after extinction
correction) and $\sigma$ = 1.3 $\pm$ 0.3. We conclude that the ground--based
dataset of Fleming \etal (1995) for NGC 4494 and our WFPC2 data {\it are}
consistent, albeit within some large scatter.  

Next we consider NGC 4565, in particular is it at the same distance as NGC 4494
(and NGC 4278) ? Figure 7 shows the Fleming \etal data for the spiral NGC 4565.
The data lie between the best fit Gaussian to our NGC 4494 data and one scaled
down by a factor of two. Thus, as with NGC 4494, the ground--based data for NGC
4565 is consistent with a Gaussian of $m_V ^0$ = 23.05. It is difficult to rule
out the possibility that the turnover magnitude is brighter by $\sim$ 0.5$^m$
(this difference is too large to be explained as metallicity effect Ashman
\etal 1995).

We now calculate the distance modulus for our data from the apparent turnover
magnitude.  As noted in the introduction, Ashman \etal (1995) have advocated
that a small correction be applied to the universal value based on GC
metallicity.  Such a metallicity correction has been shown by them to improve
GCLF distance estimates. In the absence of spectroscopic measures, we can
estimate the mean metallicity of a GC system assuming [Fe/H] = 5.051 (V--I) --
6.096 (Couture \etal 1990). Using the mean color from Forbes \etal (1996), we
derive [Fe/H] = --0.79 for NGC 4278 and --0.84 for NGC 4494, giving metallicity
corrections of $\Delta$M$_V$ = 0.16 and 0.14 respectively.  Thus we make the
universal value of M$_V ^0$ = --7.62 $\pm$ 0.08 (Sandage \& Tammann 1995)
fainter by 0.16 or 0.14 magnitudes.  

In Table 2 we summarize distance determinations from other workers for galaxies
in the Coma I cloud. We include the GCLF distance for NGC 4565 from Fleming
\etal (1995), after making a 0.05 Galactic extinction correction and assuming
M$_V ^0$ = --7.62 $\pm$ 0.08 with no metallicity correction.  From V band
surface brightness fluctuations, Simard \& Pritchet (1994) estimate (m -- M) =
30.88 $\pm$ 0.3 for NGC 4494, and 30.08 $\pm$ 0.07 for NGC 4565. Using the
planetary nebulae luminosity function (PNLF) method, Jacoby, Ciardullo \&
Harris (1996) get 30.54 $\pm$ 0.05 for NGC 4494 and 30.21 $\pm$ 0.08 for NGC
4565. From the a mass model of the Virgo region and an {\it assumption} about
the `triple value ambiguity' in velocity, Tully \& Shaya (1984) give distances
to NGC 4494 and NGC 4565 which correspond to 30.34 and 30.21 respectively.  We
will assume an error of $\pm$ 0.3$^m$ on these estimates.  Finally, Aaronson \&
Mould (1983) find (m -- M) = 30.42 $\pm$ 0.32 for several spiral galaxies in
the direction of the Coma I cloud using the infrared Tully--Fisher relation.  

The distance modulus for NGC 4494 ranges from (m -- M) = 30.34 (mass models of
Tully \& Shaya 1984) to 30.88 (SBF work of Simard \& Pritchet 1994) with a
weighted mean of 30.54 $\pm$ 0.07.  Our value for NGC 4278 (30.61 $\pm$ 0.14)
is consistent with the NGC 4494 mean.  Most recently, I band SBF measurements
also indicate that NGC 4278 and NGC 4494 have essentially the same distance
modulus, albeit at the upper range of values listed in Table 2 (Tonry 1996). 
For NGC 4565 the weighted mean distance modulus is 30.10 $\pm$ 0.05. This is
0.4$^m$ or 2 Mpc closer than NGC 4494. This suggests that NGC 4565 lies in the
foreground relative to NGC 4494.  

It has been suggested that the Coma I cloud may consist of two sub--groups, one
around NGC 4565 and the other associated with the S0 galaxy NGC 4274 and NGC
4278 (de Vaucouleurs 1976).  This claim was questioned by Gregory \& Thompson
(1977) who found no particular evidence that the Coma I cloud formed two
sub--groups. However, they did note that the group formed a bar--like structure
of dimensions 0.9 $\times$ 2.5 Mpc and that the galaxies with the {\it lower}
velocities tend to be systematically fainter, i.e. located further from us. 
Table 2 shows that in each case, other workers found NGC 4494 to be more
distant than NGC 4565 for a given distance method.  Our preferred
interpretation is that both NGC 4494 and NGC 4278 are at the same distance and
are located at the far end of the Coma I cloud, whereas NGC 4565 is located 2
Mpc closer at the front end of the group.  

Aaronson \& Mould (1983) caution that the redshifts for galaxies in the Coma I
cloud may not correlate well with distance, given the proximity of the group to
the Virgo cluster. Nevertheless, we will attempt to estimate the Hubble
constant from our measurements.  If we calculate the mean velocity of the eight
galaxies associated with NGC 4274 sub--group (de Vaucouleurs 1976), and include
NGC 4494, we get 880 km s$^{-1}$.  The correction for motion with respect to
the Local Group using solution number 2 of Yahil, Tammann \& Sandage (1977) is
$\sim$ --40 km s$^{-1}$, giving 840 km s$^{-1}$. The Virgocentric infall
component from Tammann \& Sandage (1985) is $\sim$ 200 km s$^{-1}$, which gives
a corrected recession velocity of 1020 km s$^{-1}$. Using a distance modulus of
30.54 and this corrected velocity we estimate a Hubble constant of $\sim$ 80 km
s$^{-1}$ Mpc$^{-1}$.  

Finally, we have derived the local (within 100$^{''}$ radius) and total GC
specific frequency ($S$) for each galaxy following the method described in
Forbes (1996).  These give similar results of $\sim$ 5 for NGC 4278 and $\sim$
2 for NGC 4494.  The richness of the GC system around NGC 4494 appears to lower
than that of a typical elliptical ($S$ $\sim$ 5; van den Bergh 1995).  

\section{Conclusions}

Using WFPC2 data of Forbes \etal (1996) we have fit the globular cluster
luminosity function (GCLF) of two ellipticals, NGC 4278 and NGC 4494. The first
of which is generally thought to lie in the Coma I cloud, whereas the latter
has been suggested to lie in the background.  Both the Gaussian and $t_5$
profile fits give similar results, namely a turnover magnitude of m$_V^0$ =
23.23 $\pm$ 0.11 for NGC 4278 and m$_V^0$ = 23.07 $\pm$ 0.13 for NGC 4494. The
fitted dispersions ($\sigma$ $\sim$ 1.1$^m$) are somewhat smaller than typical
values for other ellipticals.  The limiting magnitude, as determined by
completeness tests, is about 2 magnitudes fainter than these values.  We derive
distance modulii of 30.61 $\pm$ 0.14 and 30.50 $\pm$ 0.15 for NGC 4278 and NGC
4494 respectively, assuming an absolute turnover magnitude of M$_V$ = --7.62
$\pm$ 0.08 from Sandage \& Tammann (1995) and a small metallicity correction
based on the precepts of Ashman \etal (1995).  We compare our distance measure
with the ground--based GCLF study of Fleming \etal (1995) and other distance
determinations for galaxies in the Coma I cloud. Our distance modulii lie
within the range of published values. We conclude that both NGC 4278 and NGC
4494 {\it are} members of the Coma I cloud, and speculate that they lie at the
far end of a bar structure; the near end of which is associated with NGC 4565. 
Finally, we make a rough estimate of the Hubble constant and globular cluster
specific frequency from our data.  

\noindent
{\bf Acknowledgments}\\
We are particularly grateful to J. Secker for the use of his maximum likelihood
code and useful suggestions. We also thank J. Blakeslee, J. Brodie and C.
Grillmair for helpful discussions. The referee is thanked for several
suggestions that have improved the paper. This research was funded by the HST
grant AR-05794.01-94A\\

\noindent{\bf References}

\noindent
Aaronson, M., \& Mould, J. 1983, ApJ, 265, 1\\
Ashman, K. M., Conti, A., \& Zepf, S. E. 1995, AJ, 110, 1164\\ 
Blakeslee, J. P., \& Tonry, J. L. 1996, ApJ, in press\\
Burrows, C., \etal 1993, Hubble Space Telescope Wide Field and
Planetary Camera 2 Instrument Handbook, STScI\\
Carollo, C. M., Franx, M., Illingworth, G. D., \& Forbes, D. A. 1996,
ApJ, submitted\\
Couture, J., Harris, W. E., \& Allwright, J. W. B., 1990, ApJS, 73, 671\\
de Vaucouleurs, G. 1976, Stars and Stellar Systems, edited by A. Sandage,
M. Sandage and J. Kristian (Chicago: University of Chicago Press) v9, p557\\ 
Forbes, D. A. 1996, AJ, in press\\
Forbes, D. A., Elson, R. A. W., Phillips, A. C., 
Illingworth, G. D. \& Koo, D. C. 1994, ApJ, 437, L17\\
Forbes, D. A., Franx, M., Illingworth, G. D., \& Carollo, C. M. 1996,
ApJ, in press\\
Fleming, D. E. B., Harris, W. E., Pritchet, C. J., \& Hanes,
D. A. 1995, AJ, 109, 1044\\
Gregory, S. A., \& Thompson, L. A. 1977, ApJ, 213, 345\\ 
Hanes, D. A. 1977, MNRAS, 180, 309\\
Jacoby, G. H., \etal 1992, PASP, 104, 599\\
Jacoby, G. H., Ciardullo, R., \& Harris, W. E. 1996, ApJ, 462, 1\\
Sandage, A., \& Tammann, G. A. 1995, ApJ, 446, 1\\
Secker, J., \& Harris, W. E. 1993, AJ, 105, 1358\\
Simard, L., \& Pritchet, C. J. 1994, AJ, 107, 503\\
Stetson, P. B., 1987, PASP, 99, 191\\
Tammann, G. A., \& Sandage, A. 1985, ApJ, 294, 81\\
Tonry, J. L. 1996, The Extragalactic Distance Scale workshop held at
Space Telescope Science Institute\\
Tully, R. B., \& Shaya, E. J. 1984, ApJ, 281, 31\\
van den Bergh, S. 1995, AJ, 110, 2700\\
Yahil, A., Tammann, G. A., \& Sandage, A. 1977, ApJ, 217, 903\\

\begin{figure*}[p]
\centerline{\psfig{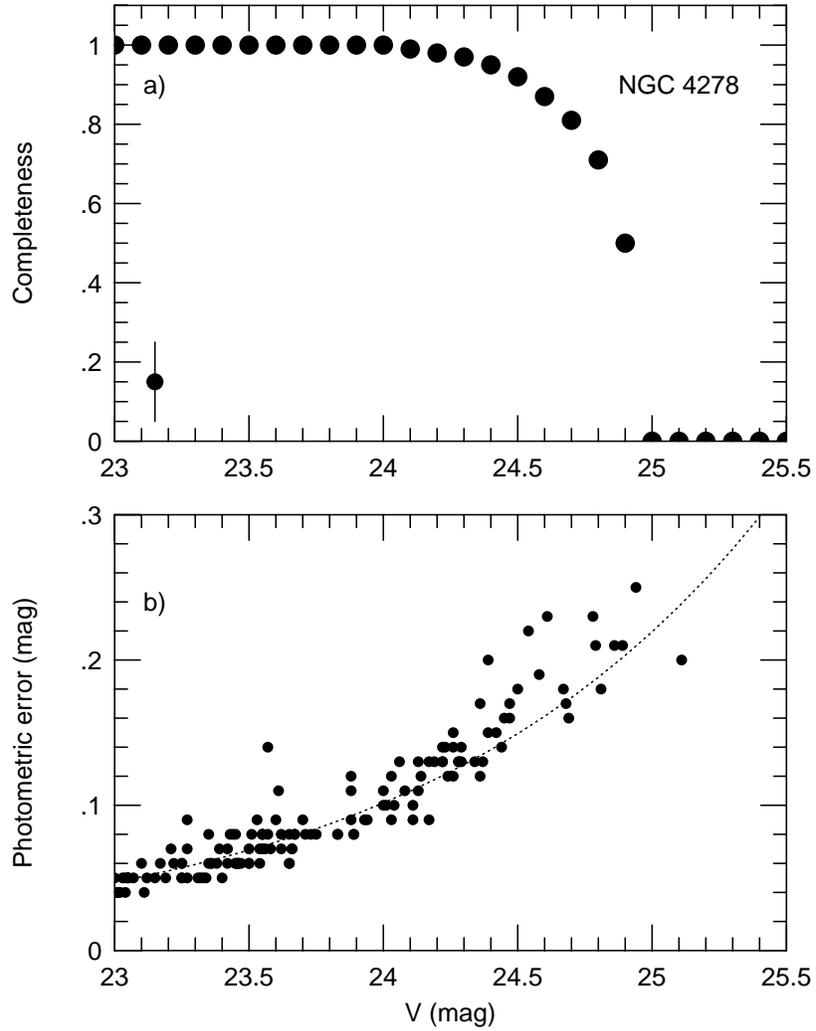}}
\caption{\label{fig1}
Completeness function for NGC 4278 from simulations. Circles show the fraction
of simulated GCs detected in 0.1 magnitude bins. A typical error bar is shown
in the lower left.  {\bf b)} Photometric error for NGC 4278 as a function of GC
V magnitude determined from DAOPHOT. Circles show the data points, and the
dashed line an exponential fit to the data of the form p.e. = exp~[a~(V~--~b)].
}
\end{figure*}

\begin{figure*}[p]
\centerline{\psfig{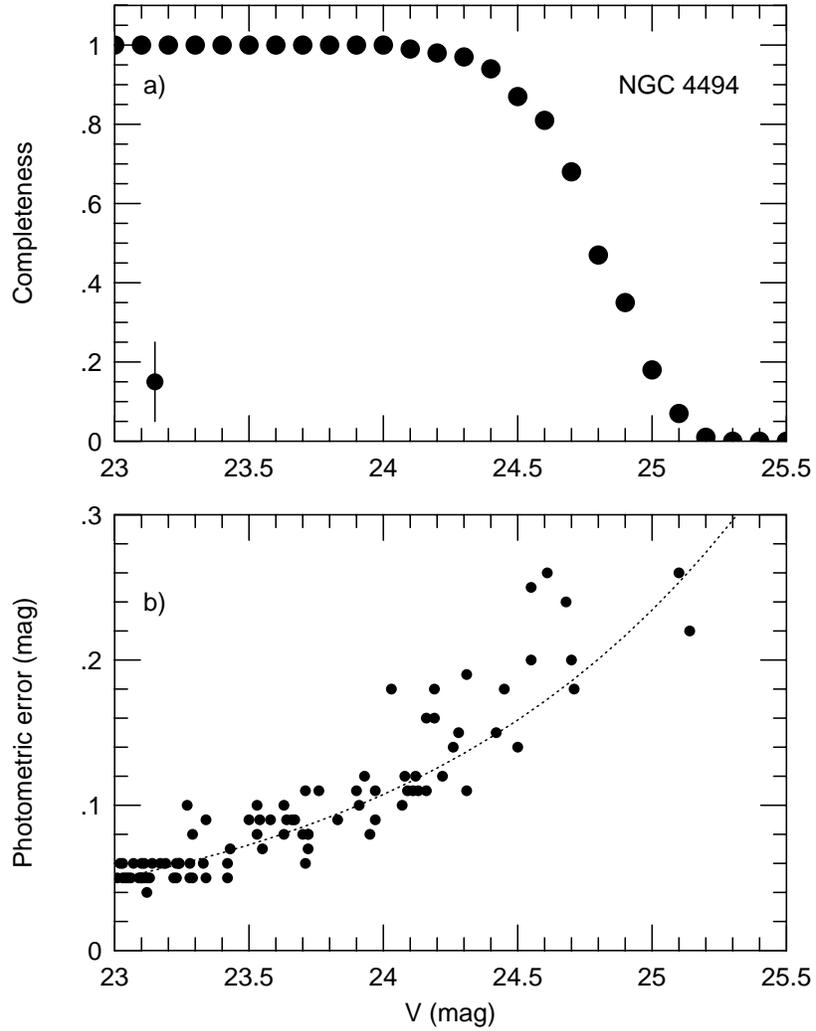}}
\caption{\label{fig2}
Completeness function for NGC 4494 from simulations. Circles show the fraction
of simulated GCs detected in 0.1 magnitude bins. A typical error bar is shown
in the lower left.  {\bf b)} Photometric error for NGC 4494 as a function of GC
V magnitude determined from DAOPHOT. Circles show the data points, and the
dashed line an exponential fit to the data of the form p.e. = exp~[a~(V~--~b)].
}
\end{figure*}

\begin{figure*}[p]
\centerline{\psfig{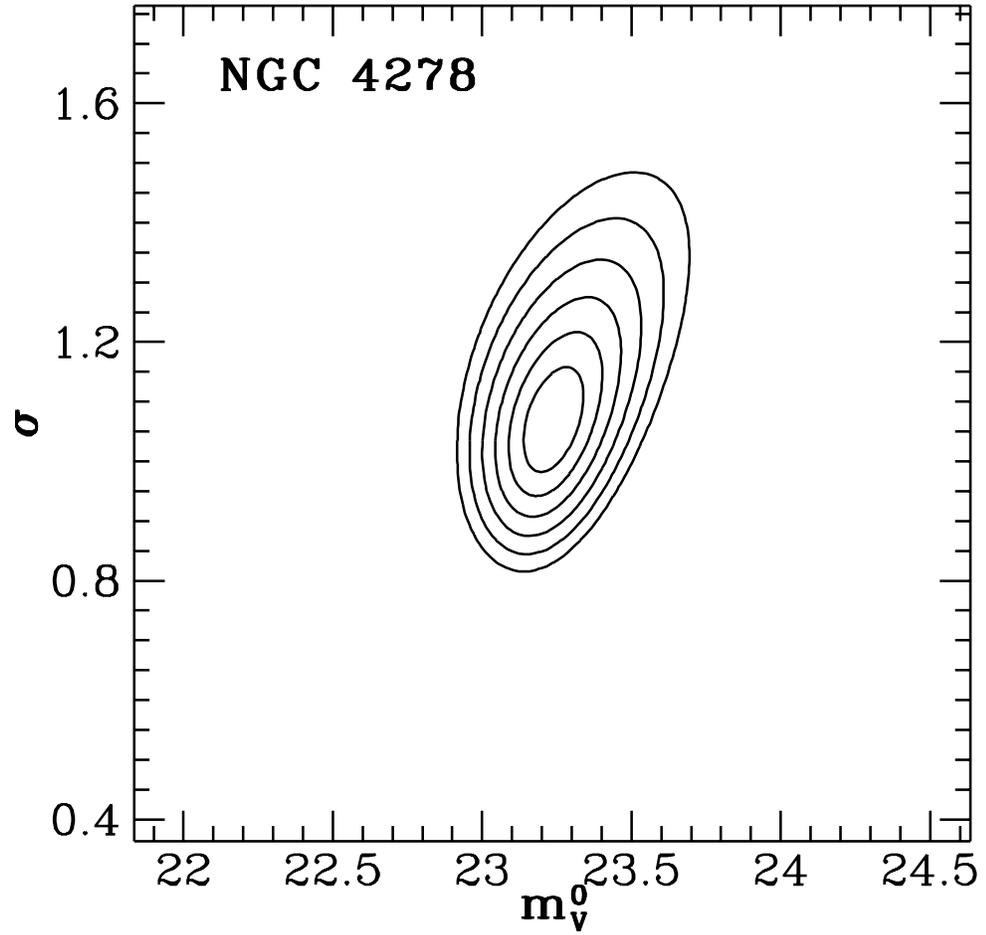}}
\caption{\label{fig3}
Probability contours for a Gaussian fit to the globular cluster luminosity
function of NGC 4278.  Contours represent 0.5 to 3 standard deviations
probability limits from the best estimate (see Table 1).
}
\end{figure*}

\begin{figure*}[p]
\centerline{\psfig{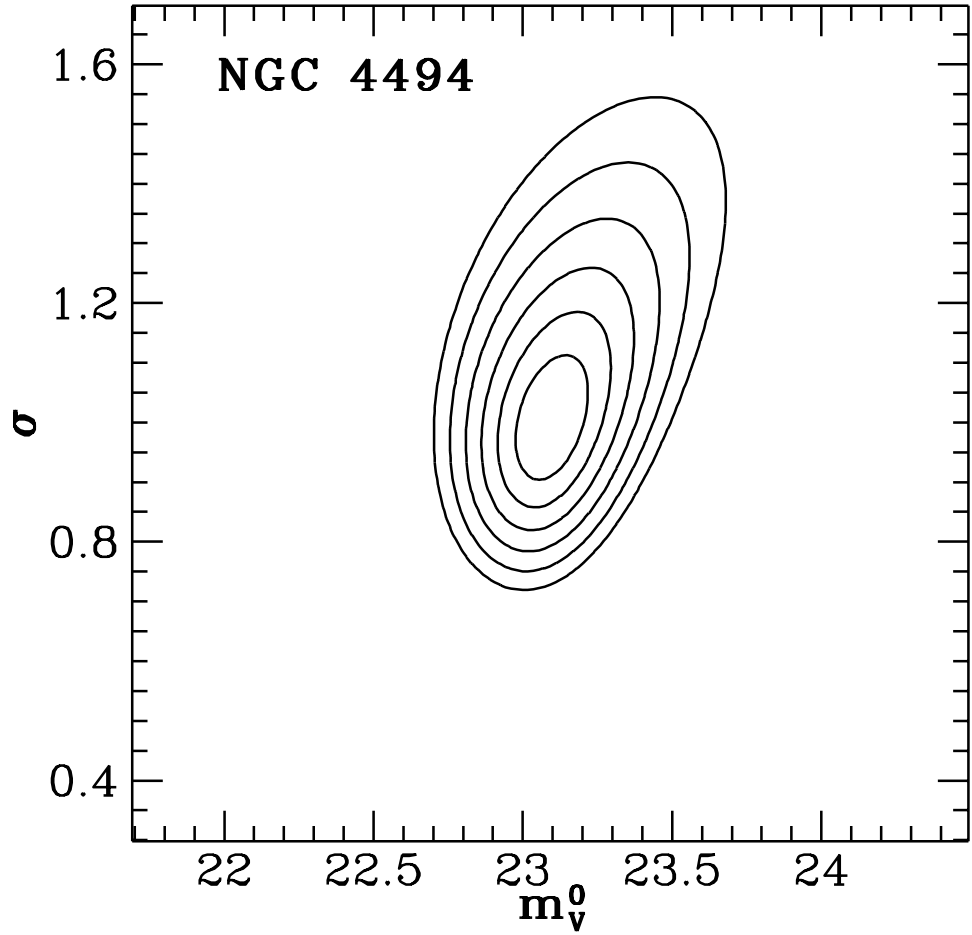}}
\caption{\label{fig4}
Probability contours for a Gaussian fit to the globular cluster luminosity
function of NGC 4494.  Contours represent 0.5 to 3 standard deviations
probability limits from the best estimate (see Table 1).
}
\end{figure*}

\begin{figure*}[p]
\centerline{\psfig{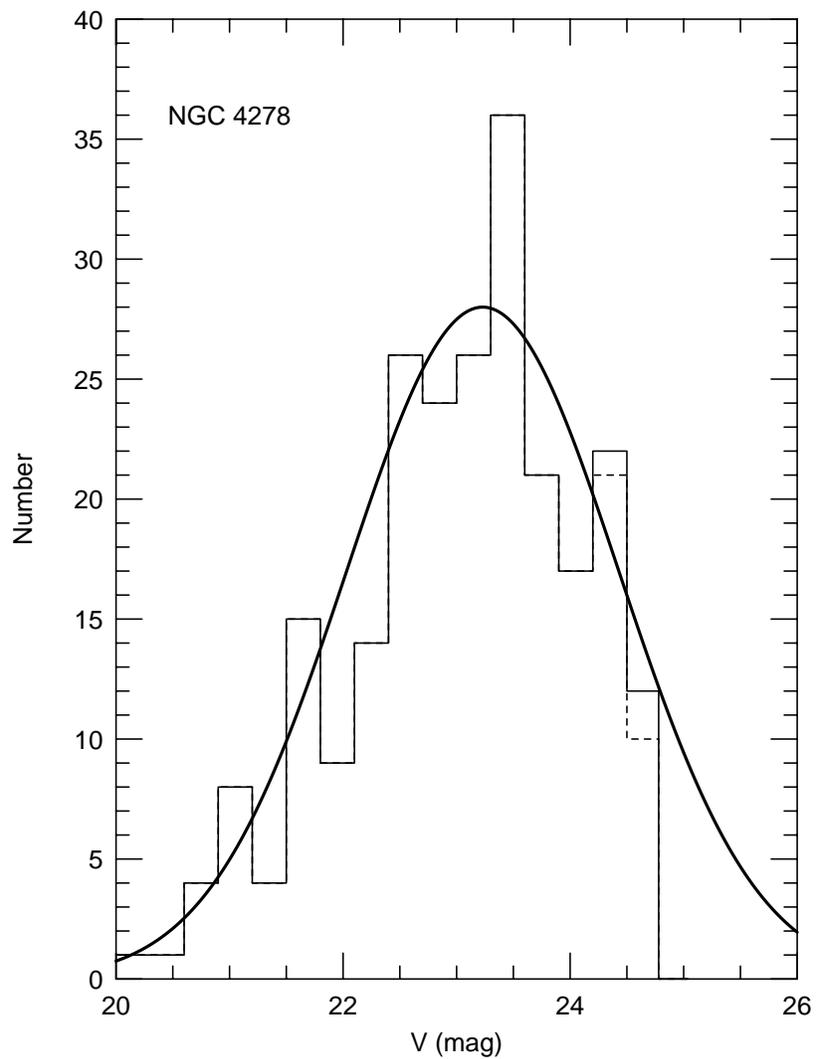}}
\caption{\label{fig5}
Globular cluster luminosity function for NGC 4278. The raw data is shown by a
dashed line, and by a thin solid line after completeness correction has been
applied. The maximum likelihood best fit Gaussian profile, which includes the
effects of photometric error and background contamination, is superposed as a
thick solid line.  Note that the fitting procedure does not use binned data.  
}
\end{figure*}

\begin{figure*}[p]
\centerline{\psfig{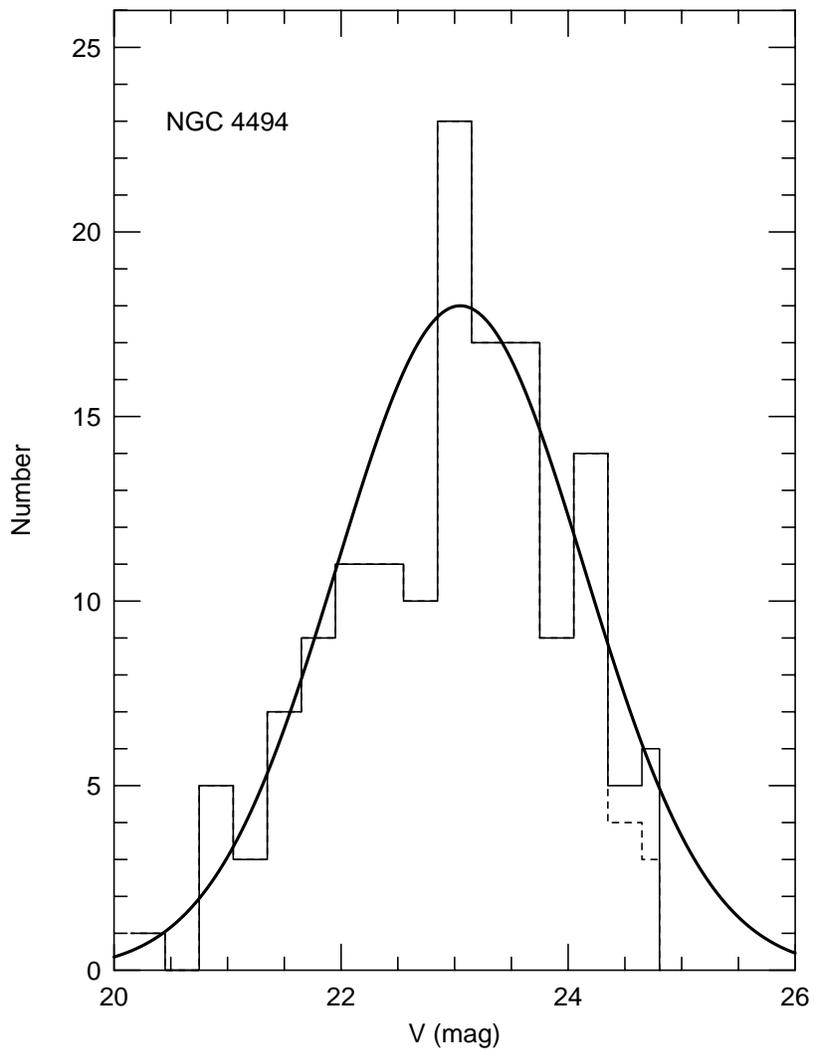}}
\caption{\label{fig6}
Globular cluster luminosity function for NGC 4494. The raw data is shown by a
dashed line, and by a thin solid line after completeness correction has been
applied. The maximum likelihood best fit Gaussian profile, which includes the
effects of photometric error and background contamination, is superposed as a
thick solid line.  Note that the fitting procedure does not use binned data.  
}
\end{figure*}

\begin{figure*}[p]
\centerline{\psfig{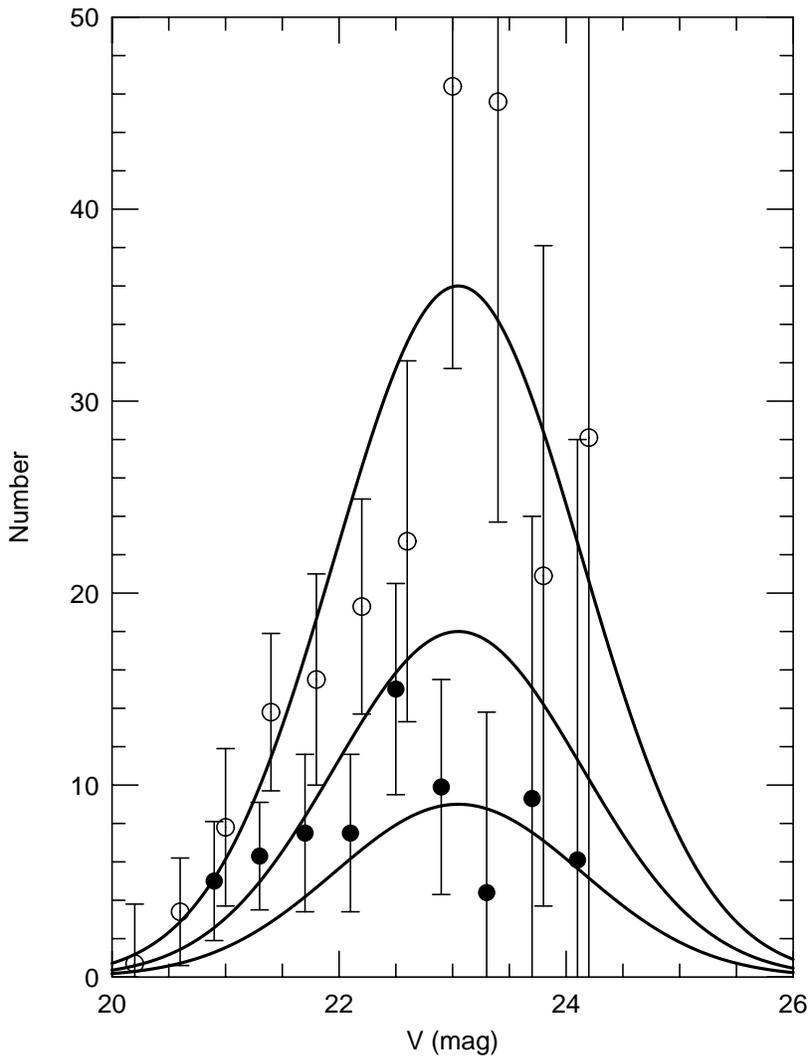}}
\caption{\label{fig7}
Comparison of the ground--based data from Fleming \etal (1995) with our data
for NGC 4494. Open circles represent the Fleming \etal data for NGC 4494, and
the filled circles for NGC 4565.  The three solid lines represent our best fit
Gaussian to the WFPC2 data on NGC 4494, and scaled up and down arbitrarily by a
factor of two (the turnover magnitude is held fixed but the dispersion is
allowed to vary).  Within the large scatter, the ground--based data for NGC
4494 and NGC 4565 could be consistent with a turnover magnitude of $m_V ^0$ =
23.05, as found from the WFPC2 data.  
}
\end{figure*}

\begin{figure*}
\centerline{\psfig{figure=table1.epsi,width=300pt}}
\end{figure*}

\begin{figure*}
\centerline{\psfig{figure=table2.epsi,width=300pt}}
\end{figure*}

\end{document}